\documentclass[twoside,12pt]{article}
\usepackage{CJK}
\usepackage{indentfirst}
\usepackage{bm}
\usepackage{color}
\usepackage{epsfig}
\usepackage{mdsymbol}
\usepackage{graphicx}
\usepackage{epstopdf}
\usepackage{cite}
\usepackage{booktabs}

\begin{document}

\begin{CJK*}{GBK}{song}

\thispagestyle{empty} \vspace*{0.8cm}\hbox
to\textwidth{\vbox{\hfill\huge\sf Commun. Theor. Phys.\hfill}}
\par\noindent\rule[3mm]{\textwidth}{0.2pt}\hspace*{-\textwidth}\noindent
\rule[2.5mm]{\textwidth}{0.2pt}


\begin{center}
\LARGE\bf A One-Dimensional Discrete Boltzmann Method for Multidimensional Compressible Flows
\end{center}

\footnotetext{\hspace*{-.45cm}\footnotesize $^\dag$Chuandong Lin, E-mail: linchd3@mail.sysu.edu.cn }

\begin{center}
\rm Yaofeng Li\textsuperscript{1},Chuandong Lin\textsuperscript{1,2,3}${^\dagger}$
\end{center}

\begin{center}
\begin{footnotesize} \sl
\textsuperscript{1}Sino-French Institute of Nuclear Engineering and Technology, Sun Yat-sen University, Zhuhai 519082, China.
\end{footnotesize}

\begin{footnotesize} \sl
	\textsuperscript{2}State Key Laboratory of Explosion Science and Safety Protection, Beijing Institute of Technology, Beijing 100081, China.
\end{footnotesize}

\begin{footnotesize} \sl
	\textsuperscript{3}Key Laboratory for Thermal Science and Power Engineering of Ministry of Education, Department of Energy and Power Engineering, Tsinghua University, Beijing 100084, China.
\end{footnotesize}

\end{center}

\begin{center}
\footnotesize (Received XXXX; revised manuscript received XXXX)

\end{center}

\vspace*{2mm}

\begin{center}
\begin{minipage}{15.5cm}
\parindent 20pt\footnotesize

\end{minipage}
\end{center}

\begin{center}
	\begin{minipage}{15.5cm}
		\parindent 20pt\footnotesize
A simple and efficient one-dimensional discrete Boltzmann method is developed for compressible flows with tunable specific heat ratios by incorporating extra degrees of freedom. To guarantee Galilean invariance in numerical simulations, a discrete velocity set is constructed with high spatial symmetry. Furthermore, an operator-splitting scheme is proposed to extend the one-dimensional kinetic formulation to simulations of one-, two-, and three-dimensional flow systems within a unified framework. The proposed model and numerical method are verified and validated against several benchmark problems, including the Sod shock tube, Lax shock tube, 2D Riemann problem, uniform translational flow, and acoustic wave propagation. The results demonstrate the accuracy, robustness, and flexibility of the present approach for compressible flow simulations.
    \end{minipage}
\end{center}

\begin{center}
\begin{minipage}{15.5cm}
\begin{minipage}[t]{2.3cm}{\bf Keywords:}\end{minipage}
\begin{minipage}[t]{13.1cm}
Discrete Boltzmann method, operator splitting method, compressible flow.
\end{minipage}\par\vglue8pt
\end{minipage}
\end{center}

\vspace*{5mm}

\section{Introduction}

Mesoscopic methods provide a powerful framework that bridges the gap between microscopic molecular dynamics and macroscopic continuum mechanics \cite{ChenSY1998ARFM, Succi2001, Xu2022}. While microscopic models resolve the dynamics of particles and thus incur prohibitively high computational costs for large systems, macroscopic models describe the evolution of averaged quantities such as density, velocity, and pressure with limited ability to capture nonequilibrium effects. In contrast, mesoscopic kinetic models, typically based on the Boltzmann or Enskog equations, evolve distribution functions that retain essential kinetic information while maintaining computational efficiency. These models therefore provide a physically rigorous yet numerically tractable bridge between the microscopic and macroscopic descriptions of fluid behavior. Among various mesoscopic approaches, the lattice Boltzmann method (LBM) \cite{Guo2013, Succi2001} and the discrete Boltzmann method (DBM) \cite{Xu2022, Xu2024FOP} have emerged as two representative schemes capable of efficiently simulating complex nonequilibrium and multiscale flow phenomena.

Since the advent of lattice-gas or cellular automaton methods, the LBM has generally been classified into two types: one for constructing coarse-grained physical models \cite{Mccoy2009, Lee1952PR, Succi2001}, and the other for numerically solving governing equations \cite{He2009, Guo2013, Huang2015}. The latter dominates the literature, so LBM typically refers to this category and is often called the standard LBM. Nevertheless, LBM can also be employed for the construction of cross-scale physical models, commonly referred to as the DBM. These two types differ in their objectives and construction principles, operate in complementary dimensions, and are each well justified in their respective contexts \cite{Xu2024FOP}. Unlike the standard LBM, which primarily serves as a numerical tool, the DBM is grounded in nonequilibrium statistical physics and is capable of capturing both hydrodynamic and thermodynamic nonequilibrium effects. It ensures physical consistency by constructing discrete distribution functions based on kinetic theory and decouples the time step, spatial resolution, and discrete velocity sets, offering greater flexibility and accuracy in modelling complex fluid phenomena. These factors have enabled the DBM to be widely applied over the past decade to multiphase flows \cite{GanYB2022JFM, WangSE2023CP}, combustion \cite{HuangWH2025ATE, WuQB2025CF}, plasma \cite{LiuZP2023JMES, SongJH2024POF}, Riemann problems \cite{ GuoQH2025POF}, sound wave propagation \cite{LinCD2019PRE}, and various fluid instabilities \cite{ChenF2018POF, LaiHL2024CF, LiYF2022FOP, LaiHL2025POF, LiYF2024POF, LinCD2025CMA}. 

Conventionally, the one-dimensional (1D) formulation is used for 1D problems, the two-dimensional (2D) formulation can be employed for both 1D and 2D configurations, and the three-dimensional (3D) formulation is applicable to 1D-3D cases. A natural question then arises: can the 1D model also be extended to simulate higher-dimensional systems? The answer is affirmative. In this context, we propose a preliminary cross-dimensional DBM framework capable of simulating 1D problems and readily extending to 2D and 3D cases. To achieve this, the operator splitting method \cite{Marchuk1968AM} is adopted, which decomposes a complex evolution problem into a sequence of simpler subproblems solved successively to approximate the full solution. Splitting schemes can be first-order, such as Godunov splitting \cite{Godunov1959MS}, or second-order, such as Strang splitting \cite{Strang1968SIAM}, with higher-order errors analyzable via the Taylor-Lie series \cite{LeVeque2002}. Owing to its simplicity and modularity, operator splitting has been applied in mesoscopic simulations \cite{Dellar2011PRE, Yan2013FOP, LinCD2014PRE, Hajabdollahi2018PRE}. Subsequently, we employ first-order operator splitting to decompose the 3D Boltzmann equation into a sequence of 1D evolution processes, each advancing the system along a single spatial direction. The remainder of this paper is organized as follows: Part 2 introduces the construction of the cross-dimensional DBM, Part 3 presents its verification through the Sod shock tube, Lax shock tube, 2D Riemann problem, translational motion, and sound wave tests, and Part 4 summarizes the main conclusions.

\section{Discrete Boltzmann method}
\label{SecII}
The Bhatnagar--Gross--Krook discrete Boltzmann equations take the form:
\begin{equation}
	\frac{{\partial {{f}_i}}}{{\partial t}} + {\boldsymbol{v}_i}\cdot{\nabla}{{f}_i} =  -\frac{1}{\tau}({{f}_i}-{{f}}^{\text{eq}}_i) \tt{.}
	\label{discrete Boltzmann equations}
\end{equation}
Here, $t$ denotes time, $\boldsymbol{v}_i$ represents the discrete velocity vector, and $f_i$ and $f_i^{\mathrm{eq}}$ are the discrete distribution function and its equilibrium counterpart, respectively. The relaxation time $\tau$ determines the rate at which $f_i$ approaches $f_i^{\mathrm{eq}}$.

The principle of physical consistency requires the following relationship,
\begin{equation}
	\iint f^{\text{eq}} \Psi \mathrm{d}\boldsymbol{v}\mathrm{d}{\eta}=\sum\limits_{i}f^{\text{eq}}_{i}{\Psi}_i\tt{,}
	\label{feq_i}
\end{equation}
where $\Psi=1, \boldsymbol{v}, \boldsymbol{v}\cdot\boldsymbol{v}+{\eta}^2, \boldsymbol{v}\boldsymbol{v}, (\boldsymbol{v}\cdot\boldsymbol{v}+{\eta}^2)\boldsymbol{v}, \dots$, and ${\Psi}_i=1, \boldsymbol{v}_i, \boldsymbol{v}_i\cdot\boldsymbol{v}_i+{\eta}_i^2, \boldsymbol{v}_i\boldsymbol{v}_i, (\boldsymbol{v}_i\cdot\boldsymbol{v}_i+{\eta}_i^2)\boldsymbol{v}_i, \dots$. Here, $\boldsymbol{v}$ and $\boldsymbol{v}_i$ are employed to describe the translational energy, while $\eta$ and $\eta_i$ are associated with vibrational and/or rotational energies. The values of $\boldsymbol{v}_i$ and $\eta_i$ are tunable to enhance the robustness and accuracy of the DBM. Specifically, the values of discrete velocities $\boldsymbol{v}_i$ are typically chosen around the flow velocity $|\boldsymbol{u}|$ and sound speed $v_s = \sqrt{\gamma T}$, where $\gamma = {(D+I+2)}/{(D+I)}$, $T$ denotes the temperature, $D$ is the spatial dimension, and $I$ represents the number of extra degrees of freedom. Meanwhile, the parameters $\eta_i$ are centered near $\bar{\eta} = \sqrt{{IT}/{m}}$, with the molar mass $m=1$, in accordance with the equipartition theorem of energy. The equilibrium distribution function is expressed by
\begin{equation}
	{f}^{\text{eq}}=\rho\big(\dfrac{1}{2\pi T}\big)^{D/2}\big(\dfrac{1}{2\pi I T}\big)^{1/2}\mathrm{\exp}\big(-\dfrac{|\boldsymbol{v}-\boldsymbol{u}|^2}{2T}-\dfrac{{\eta}^2}{2IT}\big)\tt{,}
	\label{feq}
\end{equation}
where $\rho$ is the density. 

In addition, the mass, momentum, and energy are associated with the first three groups of Eq. (\ref{feq_i}), where $f^{\text{eq}}$ and $f^{\text{eq}}_{i}$ can be replaced by $f$ and $f_{i}$, respectively. To be specific, 
\begin{equation}
	\rho=\sum\limits_{i}f_{i}\tt{,}
	\label{mass}
\end{equation}
\begin{equation}
	\boldsymbol{J}=\sum\limits_{i}f_{i}{\boldsymbol{v}}_{i}=\rho \boldsymbol{u}\tt{,}
	\label{momentum}
\end{equation}
\begin{equation}
	E=\frac{1}{2}\sum\limits_{i}f_{i}(|\boldsymbol{v}_{i}|^2+{\eta^2_i})= \frac{1}{2}(D + I)\rho T +\frac{1}{2 \rho}{|\boldsymbol{J}|}^2 \tt{
		.}
	\label{energy}
\end{equation}
Accordingly, $\boldsymbol{u}$ and $T$ can be obtained from the conserved moments as
\begin{equation}
	\boldsymbol{u}=\frac{\boldsymbol{J}}{\rho}\tt{,}
	\label{u}
\end{equation}
\begin{equation}
	T= \frac{2E - \rho|\boldsymbol{u}|^{2}}{(D+I)\rho} \tt{.}
	\label{T}
\end{equation}

Furthermore, the Chapman--Enskog multiscale analysis demonstrates that the Euler equations can be recovered in the continuum limit if the first five elements of $\Psi$ and ${\Psi}_i$ are satisfied in Eq. (\ref{feq_i}). Specifically, the Euler equations read
\begin{eqnarray} 
	\frac{\partial \rho }{\partial t}+{\nabla}\cdot({\rho\boldsymbol{u}})=0\tt{,}
	\label{mass}
\end{eqnarray}
\begin{eqnarray}
	\frac{\partial ({\rho\boldsymbol{u}})}{\partial t}+{\nabla}\cdot({\rho\boldsymbol{u}}\boldsymbol{u}+p \boldsymbol{I})=0\tt{,}
	\label{momentum}
\end{eqnarray}
\begin{eqnarray}	
	\frac{\partial E }{\partial t}+{\nabla}\cdot({E \boldsymbol{u}}+p \boldsymbol{u})=0\tt{,}
	\label{energy}
\end{eqnarray}
where $p = \rho T$ denotes the pressure, and $\boldsymbol{I}$ is an identity matrix.

\begin{figure}
	\begin{center}
		\includegraphics[width=0.5\textwidth]{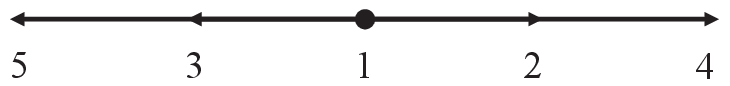}
	\end{center}
	\caption{Discrete velocity set.}
	\label{Fig01}
\end{figure}

At the Euler level, the 1D, 2D, and 3D models require five, nine, and fourteen discrete velocities, respectively \cite{LinCD2022AAS, LinCD2023CPB}. The motivation of this study is to demonstrate the idea that a 1D model is capable of simulating not only 1D physical systems but also 2D or 3D ones. For this sake, we construct the discrete velocity model D1V5 incorporating additional degrees of freedom, as illustrated in Fig. \ref{Fig01}. The discrete parameter sets are given by 
\begin{equation}
    \left\{
    \begin{aligned}
        (v_1, v_2, v_3, v_4, v_5) &= (0, v_a, -v_a, v_b, -v_b), \\
        (\eta_1, \eta_2, \eta_3, \eta_4, \eta_5) &= (\eta_a, \eta_b, \eta_b, \eta_c, \eta_c).
    \end{aligned}
    \right.
\end{equation}
The spatial derivatives are discretized using a second-order non-oscillatory, parameter-free dissipative scheme \cite{ZhangHX1991AAM}, while the temporal evolution is integrated with a first-order forward Euler scheme. 

\begin{figure}
	\begin{center}
		\includegraphics[width=0.5\textwidth]{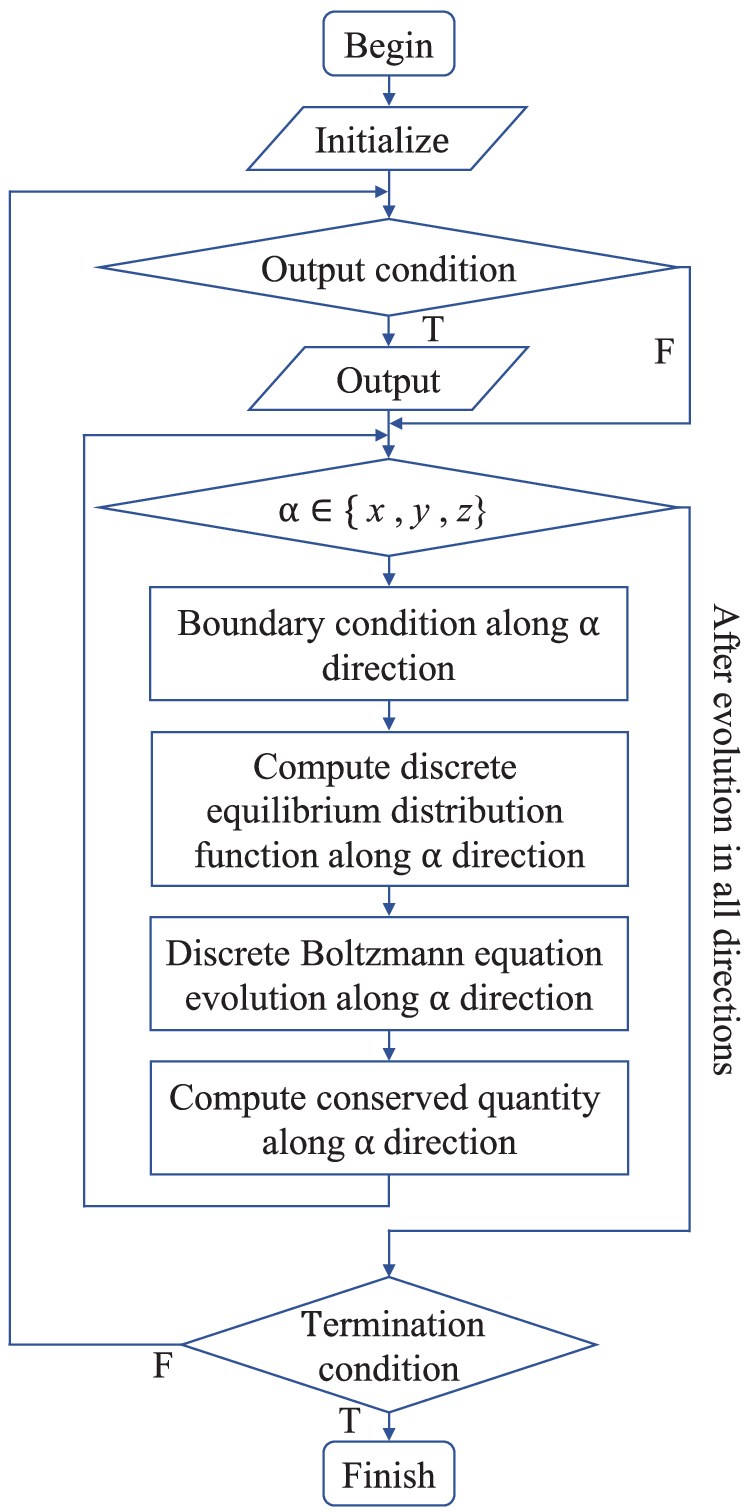}
	\end{center}
	\caption{Flowchart of the cross-dimensional DBM.}
	
	\label{Fig02}
\end{figure}

Let us introduce the flowchart of the 1D DBM applied to a 3D fluid system, as shown in Fig. \ref{Fig02}. At the initial time, the physical field is specified by the density $\rho_0$, flow velocity $\boldsymbol{u}_0=(u_{x0},u_{y0},u_{z0})$, and temperature $T_0$. These quantities serve as the initial input to the flowchart's initialization module. The evolution proceeds through three sequential steps as follows.

\textbf{Step 1:} Evolution in the $x$ direction (${{f}_{i}}\xrightarrow{x}f_{i}^{*}$)

First of all, the physical field evolves along the $x$-axis. The update procedure includes the following substeps: (i) apply the $x$-direction boundary conditions; (ii) evaluate the local equilibrium distribution $f_i^{\text{eq}} = f_i^{\text{eq}}(\rho, u_x, T)$, and assign $f_i^{\text{eq}}$ to $f_i$, i.e., $f_i^{\text{eq}} \rightarrow f_i$; (iii) advance the discrete distribution functions from $f_i$ to $f^*_i$ according to 
\begin{equation}
	\frac{\partial f_i}{\partial t} + v_{i}\frac{\partial f_i}{\partial x} = -\frac{1}{\tau}(f_i - f_i^{\text{eq}})
	\text{,}
\end{equation}
and (iv) update the physical variables from the discrete distribution functions $f^*_i$, i.e., $\rho=\rho(f^*_i)$, $u_{x}=u_{x}(f^*_i)$, and $T=T(f^*_i)$.

\textbf{Step 2:} Evolution in the $y$ direction ($f_{i}^{*}\xrightarrow{y}f_{i}^{**}$)

Afterwards, the field evolves along the $y$-axis, following the same sequence: (i) impose the $y$-direction boundary conditions; (ii) with updated $\rho(f^*_i)$ and $T(f^*_i)$, compute $f_i^{\text{eq}*} = f_i^{\text{eq}}(\rho, u_y, T)$, and set $f_i^{\text{eq}*}$ to $f_i^*$, i.e., $f_i^{\text{eq}*} \rightarrow f_i^*$; (iii) update the discrete distribution functions from $f^*_i$ to $f^{**}_i$ through 
\begin{equation}
	\frac{\partial f^*_i}{\partial t} + v_{i}\frac{\partial f^*_i}{\partial y} = -\frac{1}{\tau}\left(f^*_i - f_i^{\text{eq}*}\right),
\end{equation}
and (iv) update the macroscopic quantities $\rho=\rho(f^{**}_i)$, $u_{y}=u_{y}(f^{**}_i)$, and $T=T(f^{**}_i)$.

\textbf{Step 3:} Evolution in the $z$ direction (${{f}^{**}_{i}}\xrightarrow{z}f_{i}^{***}$)

Finally, the field evolves along the $z$-axis: (i) apply the $z$-direction boundary conditions; (ii) with renewed $\rho(f^{**}_i)$ and $T(f^{**}_i)$, compute $f_i^{\text{eq}**} = f_i^{\text{eq}}(\rho, u_z, T)$, and assign $f_i^{\text{eq}**}$ to $f^{**}_i$, i.e., $f_i^{\text{eq}**}\rightarrow f^{**}_i$; (iii) advance the discrete distribution functions from $f^{**}_i$ to $f^{***}_i$ following 
\begin{equation}
	\frac{\partial f^{**}_i}{\partial t} + v_{i}\frac{\partial f^{**}_i}{\partial z} = -\frac{1}{\tau}\left(f^{**}_i - f_i^{\text{eq}**}\right),
\end{equation}
and (iv) update the macroscopic variables $\rho=\rho(f^{***}_i)$, $u_{z}=u_{z}(f^{***}_i)$, and $T=T(f^{***}_i)$.

At this stage, one full evolution cycle is completed, and the distribution functions have been advanced from their previous values $f_i(t)$ to the updated values $f_i(t+\Delta t)=f^{***}_i$. Subsequently, the physical field continues to evolve following the same three-step procedure. This operator splitting method provides a natural framework using a 1D model to simulate a cross-dimensional system, where 1D and 2D cases can be readily obtained by omitting the later steps.

\section{Numerical validation}
\label{SecIII}
To validate the reliability of the cross-dimensional model, five classical benchmarks are examined: the Sod shock tube, Lax shock tube, 2D Riemann problem, translational motion, and sound wave. 

\subsection{Sod shock tube}
\begin{figure*}
	\begin{center}
		\includegraphics[width=1.0\textwidth]{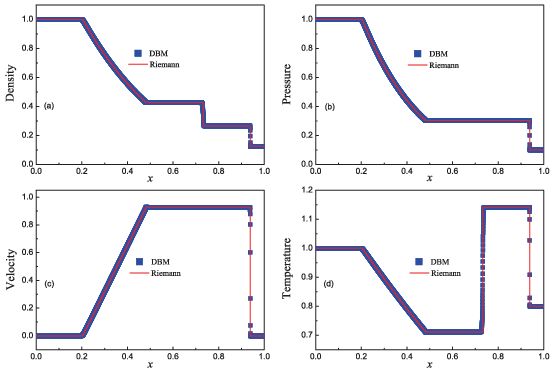}
	\end{center}
	\caption{Sod shock tube field at $t=0.25$: (a) density, (b) pressure, (c) velocity, and (d) temperature.}
	\label{Fig03}
\end{figure*}

First, the 1D Sod shock tube problem is simulated, with the initial conditions specified as 
\begin{equation}
	(\rho_\mathrm{L}, u_\mathrm{L}, T_\mathrm{L})=(1, 0, 1),\,\,\,\,\,\, (\rho_\mathrm{R}, u_\mathrm{R}, T_\mathrm{R})=(0.125, 0, 0.8)\tt{,}
	\label{Sod shock tube}
\end{equation}
where the subscripts ``$\mathrm{L}$" and ``$\mathrm{R}$" denote the left and right states separated by the initial discontinuity located at $x_0 = 0.5$. The simulations are carried out on a grid with $N_x=5000$ points, using a spatial step $\Delta x = 2 \times 10^{-4}$ and a time step $\Delta t = 5 \times 10^{-6}$. The specific heat ratio is given by $\gamma = 1.4$, and the discrete parameters are set as $(v_a, v_b, \eta_a, \eta_b, \eta_c) = (1, 5, 3.2, 0, 0)$. Free inflow and outflow boundary conditions are imposed at the left and right boundaries, respectively.

Figures \ref{Fig03} (a)-(d) illustrate the profiles of density, pressure, velocity, and temperature at $t = 0.25$. The Mach number of the shock wave is $\text{M}_a=1.66$, calculated as the ratio of the shock propagation speed to the upstream sound speed. Therefore, the flow in this configuration is compressible, as the Mach number exceeds the commonly used threshold of 0.3. Furthermore, symbols represent the DBM results, while solid lines correspond to the Riemann solutions. The numerical solution accurately resolves the characteristic wave structures, including a left-propagating rarefaction wave, a contact discontinuity in the middle, and a right-propagating shock wave. Excellent agreement is observed between the numerical and exact solutions, demonstrating that the proposed model is capable of accurately simulating compressible flows.

Moreover, a quantitative comparison of the computational efficiency of the D1V5 and D2V9 models \cite{LinCD2023CPB}  is performed. The computational time of the D1V5 is 9 s, while that of the D2V9 is 36.33 s. The computational time of the D2V9 is about 4 times that of the D1V5. This significant efficiency difference is attributed to the reduced number of discrete velocities and the lower spatial dimensionality of the evolution process: the D1V5 evolves the distribution function along a single spatial direction, whereas the D2V9 requires evolution in two spatial dimensions, resulting in a higher number of computational operations per iteration.

\subsection{Lax shock tube}
\begin{figure*}
	\begin{center}
		\includegraphics[width=1.0\textwidth]{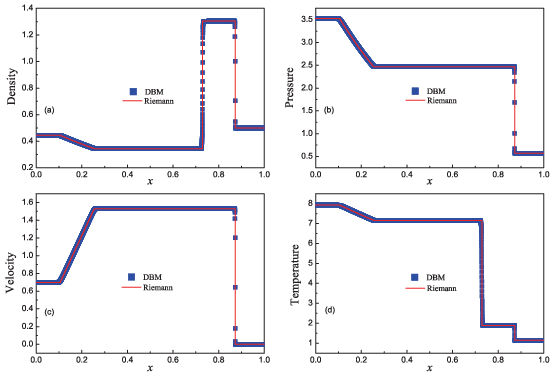}
	\end{center}
	\caption{Lax shock tube field at $t=0.15$: (a) density, (b) pressure, (c) velocity, and (d) temperature.}
	\label{Fig04}
\end{figure*}

We next consider another 1D Riemann problem, the Lax shock tube, with the initial states given by 
\begin{equation}
	(\rho_\mathrm{L}, u_\mathrm{L}, T_\mathrm{L})=(0.445, 0.698, 7.928),\,\, (\rho_\mathrm{R}, u_\mathrm{R}, T_\mathrm{R})=(0.5, 0, 1.142)\tt{.}
	\label{Lax shock tube}
\end{equation}
All other conditions are the same as those used in the Sod shock tube problem. Figures \ref{Fig04} (a)-(d) compare the DBM results with the Riemann solutions for density, pressure, velocity, and temperature at $t=0.15$, showing excellent agreement. In the rarefaction region, the density, pressure, and temperature decrease smoothly from left to right, while the velocity increases continuously. In the vicinity of the material interface, the density and temperature exhibit opposite variations, whereas the pressure and flow velocity remain constant. Meanwhile, all variables exhibit sharp jumps across the shock front which travels rightwards with a supersonic speed.

\subsection{2D Riemann problem}
\begin{figure*}
	\begin{center}
		\includegraphics[width=0.8\textwidth]{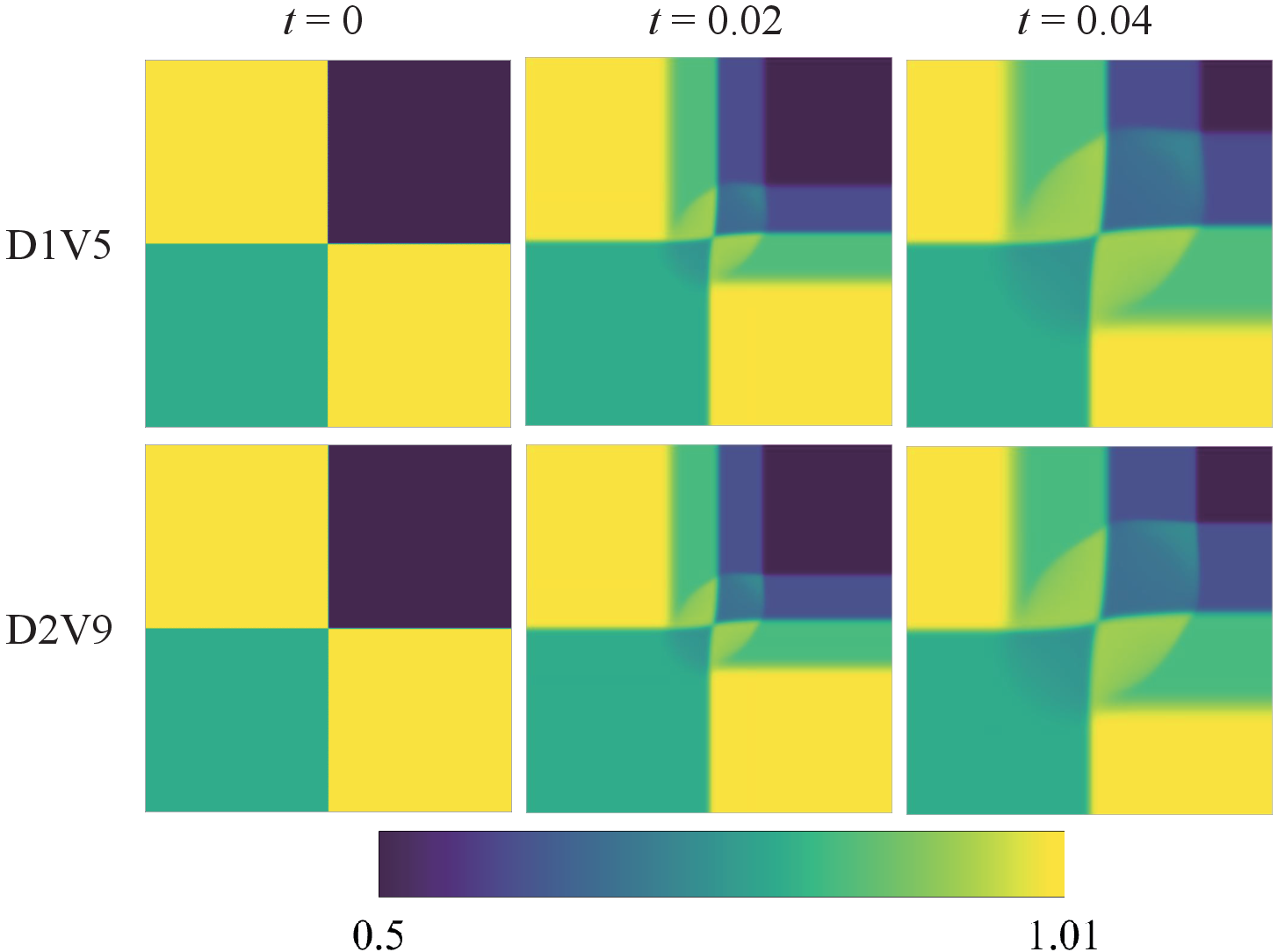}
	\end{center}
	\caption{Density contours of the 2D Riemann problem for configuration I at various time instants. Top row: D1V5; Bottom row: D2V9.}
	\label{Fig05}
\end{figure*}

\begin{figure*}
	\begin{center}
		\includegraphics[width=0.8\textwidth]{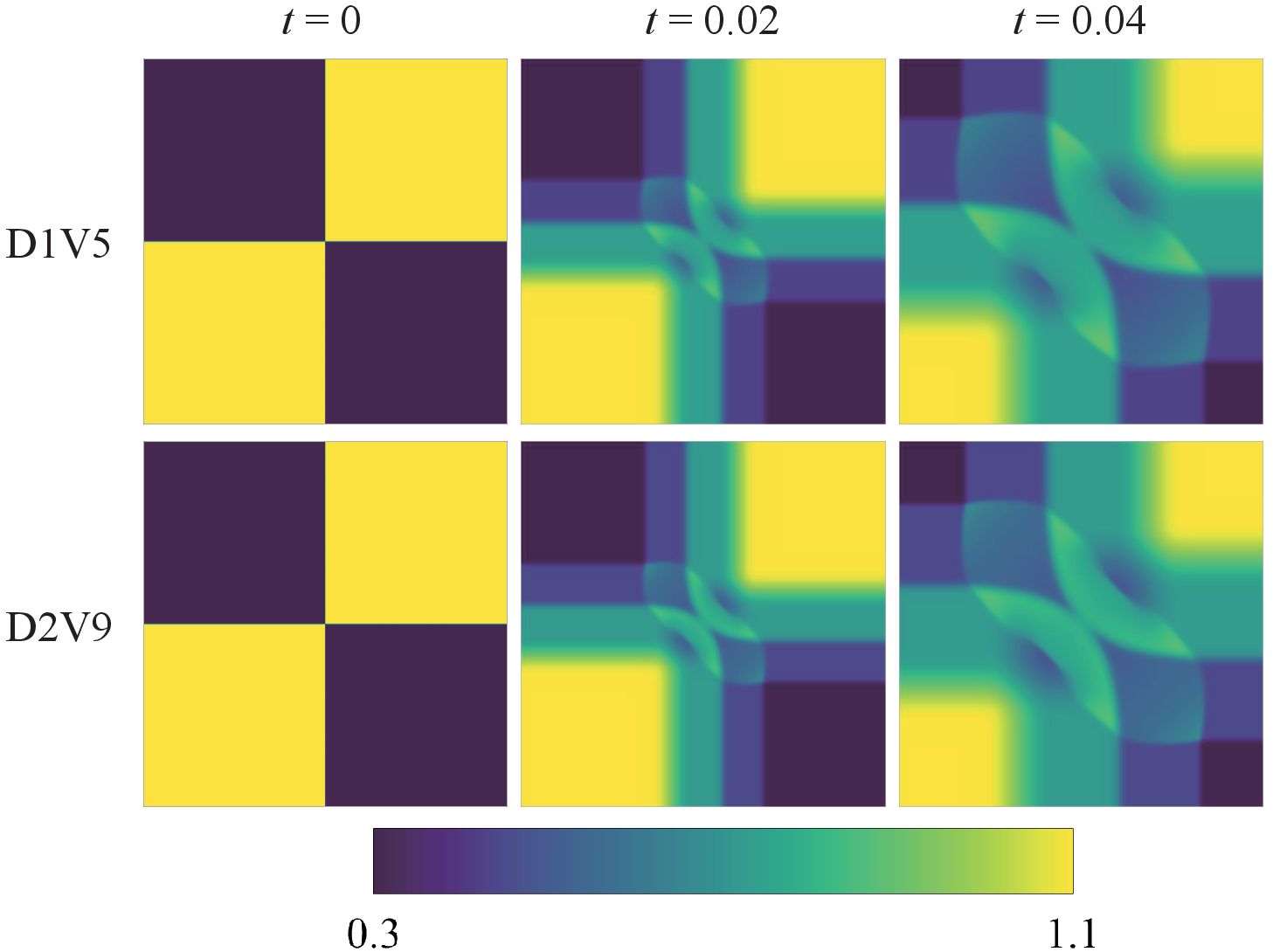}
	\end{center}
	\caption{Density contours of the 2D Riemann problem for configuration II at various time instants. Top row: D1V5; Bottom row: D2V9.}
	\label{Fig06}
\end{figure*}

This subsection utilizes the 2D Riemann problem in gas dynamics to compare the D1V5 and D2V9 models \cite{LinCD2023CPB}. The initial physical field is divided into four adjacent rectangular subdomains, each of which is assigned a uniform state for density, pressure, and flow velocity. Two representative configurations are selected for comparative analysis.
For configuration I, 
\begin{equation}
	(\rho,p,u_x,u_y)=\left\{
	\begin{aligned}
		&(0.8,1,0,0), 0< x\leqslant0.1, 0< y\leqslant0.1,\\
		&(1,1,0.05,0), 0.1< x\leqslant0.2, 0< y\leqslant0.1,\\
		&(1,1,0,0.05), 0< x\leqslant0.1, 0.1< y\leqslant0.2,\\
		&(0.5,0.6,0,0), 0.1< x\leqslant0.2, 0.1< y\leqslant0.2.\\
	\end{aligned}
	\right.
\end{equation}
For configuration II,
\begin{equation}
	(\rho,p,u_x,u_y)=\left\{
	\begin{aligned}
		&(1.1,1.1,0.05,0.05), 0<  x\leqslant0.1, 0<  y\leqslant0.1,\\
		&(0.3,0.35,0.05,0), 0.1<  x\leqslant0.2, 0<  y\leqslant0.1,\\
		&(0.3,0.35,0,0.05), 0<  x\leqslant0.1, 0.1<  y\leqslant0.2,\\
		&(1.1,1.1,0,0), 0.1<  x\leqslant0.2, 0.1< y\leqslant0.2.\\
	\end{aligned}
	\right.
\end{equation}
The simulation employs a grid of $N_x = N_y = 400$, with inflow/outflow boundary conditions applied in all directions. The grid and time steps are set to $\Delta x = \Delta y = 5 \times 10^{-4}$ and $\Delta t = 1 \times 10^{-4}$, respectively.
	
Figures \ref{Fig05} and \ref{Fig06} display the density fields for configurations I and II, respectively. Each figure shows the snapshots obtained using the D1V5 and D2V9 models at time instants $t = 0$, $0.02$, and $0.04$. For configuration I, the discrete parameters are set as $(v_a, v_b, \eta_a, \eta_b, \eta_c) = (1, 0.9, 3.8, 0, 0)$ for D1V5, and $(v_a, v_b, v_c, \eta_a, \eta_b, \eta_c) = (1, 1.9, 1, 3.8, 0, 0)$ for D2V9; For configuration II, $(v_a, v_b, \eta_a, \eta_b, \eta_c) = (1, 3.2, 1.8, 0, 0)$ for D1V5, and $(v_a, v_b, v_c, \eta_a, \eta_b, \eta_c) = (1, 3.2, 1, 1.8, 0, 0)$ for D2V9. At the interfaces between the four subdomains, various waves are generated and interact with one another, leading to the formation of complex flow structures near the central region. Meanwhile, the complex patterns simulated by the D1V5 and D2V9 models are similar to each other.

Taking configuration I as an example, a quantitative comparison of the computational efficiency of the two models is performed. The computational time of the D1V5 is 29.33 s, while that of the D2V9 is 24 s. These results indicate that, for this configuration, the computational efficiency of the D2V9 is approximately 22\% higher than that of the D1V5. This difference in efficiency results from the difference in algorithmic complexity between the two models.

\subsection{Translational motion} 
\begin{figure*}
	\begin{center}
		\includegraphics[width=1.0\textwidth]{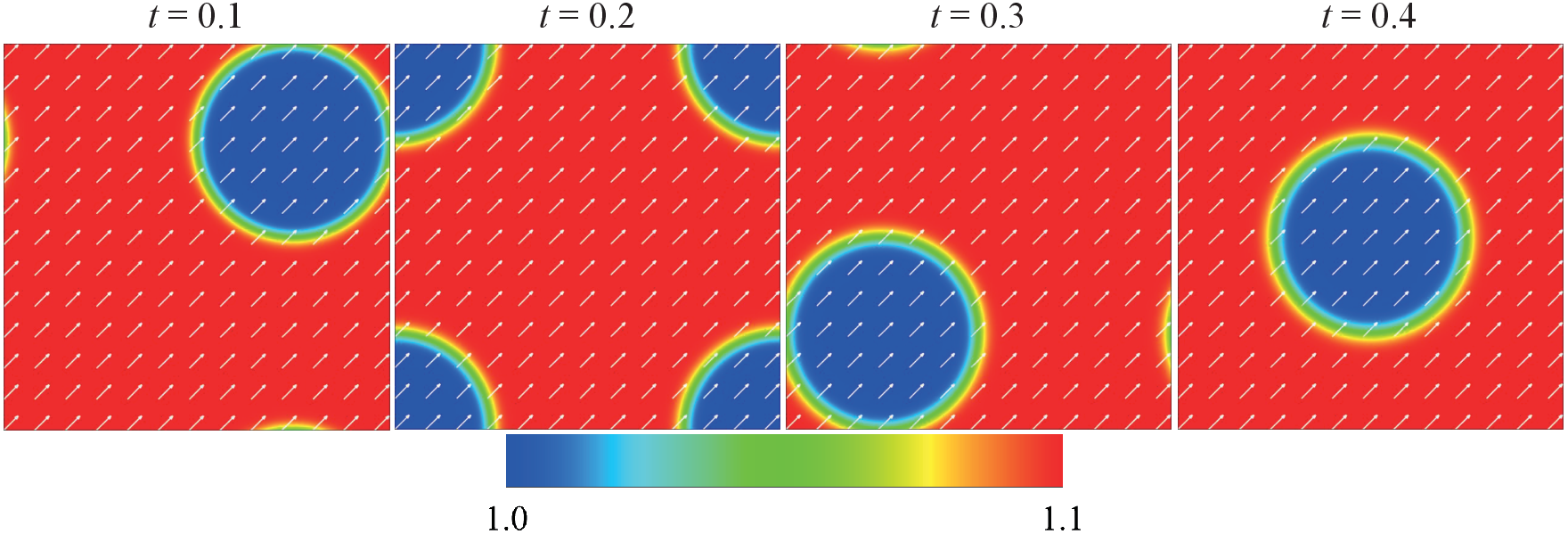}
	\end{center}
	\caption{Diagonal motion of fluid at $t = 0.1$, $0.2$, $0.3$, and $0.4$.}
	\label{Fig07}
\end{figure*}

Subsequently, Galilean invariance of the proposed model is examined through a translational motion test. A circular fluid region is initially placed at the center of the square computational domain, located at $[L_x/2, L_y/2]$, with a radius $R = L_x/4$, where $L_x = N_x \Delta x$ and $L_y = N_y \Delta y$ denote the domain width and length, respectively. The inner and outer densities are set to $\rho_{\mathrm{in}} = 1$ and $\rho_{\mathrm{out}} = 1.1$, respectively, with a smooth transition layer of thickness $W = L_x/50$ connecting the two regions. Mathematically, the initial density field is prescribed as
\begin{equation}
	\rho(y)=\frac{\rho_\mathrm{in}+\rho_\mathrm{out}}{2}-\frac{\rho_\mathrm{in}-\rho_\mathrm{out}}{2}\tanh(\frac{\sqrt{(x-\frac{L_x}{2})^2+(y-\frac{L_y}{2})^2}-R}{W})\tt{,}
	\label{Sod shock tube}
\end{equation}	
while the pressure is initially $p = 1$. The remaining parameters are specified as $N_x = N_y = 1000$, spatial steps $\Delta x = \Delta y = 2 \times 10^{-4}$, time step $\Delta t = 1 \times 10^{-5}$, specific heat ratio $\gamma = 1.5$, discrete parameters $(v_a, v_b, \eta_a, \eta_b, \eta_c) = (1, 5, 3.2, 0, 0)$. Periodic boundary conditions are applied in both spatial directions.	

A translational velocity is imposed in the diagonal direction, with $(u_x, u_y) = (0.5, 0.5)$. Snapshots of the density field at four representative times, $t = 0.1$, $0.2$, $0.3$, and $0.4$ from left to right, are shown in Fig. \ref{Fig07}. In each snapshot, the density increases from bule to red, and the flow direction is indicated by white arrows. The system is expected to translate by a distance $L_x$ at $t = 0.4$, according to the theoretical relation $u_x = L_x/t$.

\subsection{Sound wave} 
\begin{figure*}
	\begin{center}
		\includegraphics[width=1.0\textwidth]{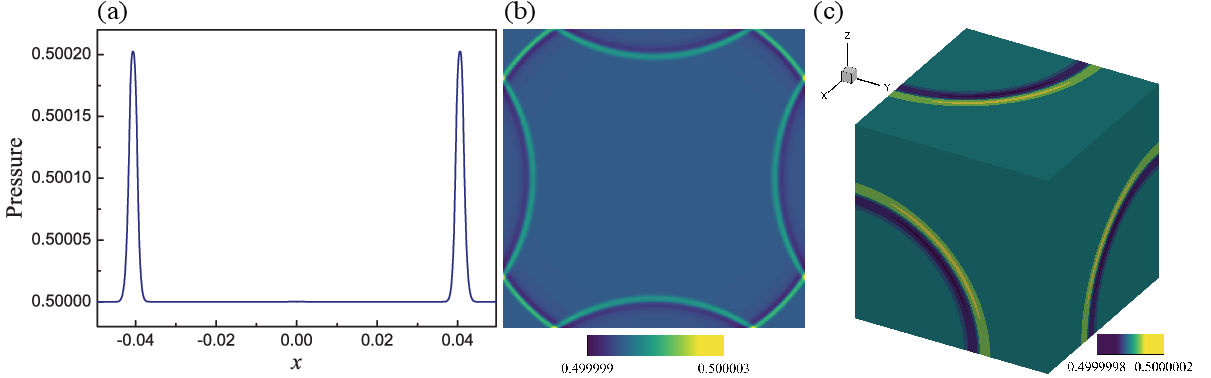}
	\end{center}
	\caption{Pressure distribution of the sound wave at $t = 0.065$: (a) 1D, (b) 2D, and (c) 3D.}
	\label{Fig08}
\end{figure*}

\begin{figure*}
	\begin{center}
		\includegraphics[width=1.0\textwidth]{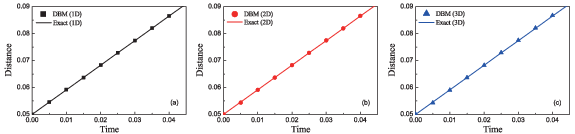}
	\end{center}
	\caption{Position of the sound wave versus time: (a) 1D, (b) 2D, and (c) 3D.}
	\label{Fig09}
\end{figure*}

Finally, the classical sound wave is employed for verification. The initial fields in 1D, 2D, and 3D cases are set to a uniform density $\rho_0 = 1$, flow velocity $\boldsymbol{u}_0 = 0$, and temperature $T_0 = 0.5$. The computational domain is discretized using $N_{\alpha} = 1000$ grids in each coordinate direction, with a spatial resolution of $\Delta \alpha = 1 \times 10^{-4}$  and a time step of $\Delta t = 1 \times 10^{-5}$. The specific heat ratio is $\gamma = 5/3$, and the discrete parameters are set as $(v_a, v_b, \eta_a, \eta_b, \eta_c) = (1, 5, 3.2, 0, 0)$. Periodic boundary conditions are applied on all boundaries. A small perturbation is imposed at the center of each computational domain. In the 1D configuration, the perturbation develops into a planar wave that propagates bidirectionally along the tube. In 2D, the disturbance expands radially as a circular wave, while in 3D it evolves into a spherical wave propagating outward from the domain center. 

Figures \ref{Fig08} (a), (b), and (c) present the sound wave pressure distributions in 1D, 2D, and 3D, respectively, at a time instant $t = 0.065$. Panel (a) shows the 1D pressure profile along the $x$-axis within the region $[-0.05, 0.05]$, where two peaks appear at approximately $x \approx \pm 0.04$ and the surrounding pressure remains close to the initial value of $0.5$. Panels (b) and (c) depict the 2D and 3D pressure fields over the domains $[-0.05, 0.05]^2$ and $[0, 0.05]^3$, respectively, where the perturbation has propagated outward from the center and already re-entered the computational domain. Figures \ref{Fig09} (a), (b), and (c) display the position of the sound wave over time in 1D, 2D, and 3D, respectively. The simulation results, obtained using the wavefront location as the reference, are consistent with the theoretical prediction $x = x_0 + v_s t$, where the sound speed is given by $v_s = \sqrt{\gamma T}$. These results demonstrate that the proposed model successfully captures the outward propagation of a small perturbation in 1D, 2D, and 3D simultaneously.

\section{Conclusions}
\label{SecIV}
In this work, we develop a 1D DBM capable of simulating not only 1D systems but also 2D and 3D configurations with an operator splitting method. Specifically, the model advances the physical fields sequentially along individual spatial directions, enabling a natural transition from one to higher-dimensional systems within a unified framework. Its performance is evaluated by the Sod shock tube, Lax shock tube, 2D Riemann problem, translational motion, and sound waves, and the results demonstrate the feasibility of the proposed approach for cross-dimensional simulations. One shortcoming of the method is the lack of the viscous, heat conductive, and other thermodynamic nonequilibrium effects when simulating 2D or 3D systems, thereby opening a promising avenue for future developments in mesoscopic fluid dynamics.

\section*{Acknowledgements}
This work is supported by National Natural Science Foundation of China (under Grant No. 12572341), Guangdong Basic and Applied Basic Research Foundation (under Grant No. 2024A1515010927), and Humanities and Social Science Foundation of the Ministry of Education in China (under Grant No. 24YJCZH163). This paper is supported by the opening project of State Key Laboratory of Explosion Science and Safety Protection (Beijing Institute of Technology). The opening project number is KFJJ26-17M. 
	
\newpage
\vspace*{-1mm}
\begin{small}\baselineskip=10pt\itemsep-2pt
\bibliography{References}	

@article{ChenSY1998ARFM,
  title={{Lattice Boltzmann method for fluid flows}},
  author={Chen, S. Y. and Doolen, G. D.},
  journal={Annu. Rev. Fluid Mech.},
  volume={30},
  number={1},
  pages={329-364},
  year={1998}
}

@book{Guo2013,
  title={{Lattice Boltzmann method and its application in engineering}},
  author={Guo, Z. L. and Shu, C.},
  year={2013},
  publisher={World Scientific},
  address   = {Singapore}
}

@book{Succi2001,
  title={{The lattice Boltzmann equation: for fluid dynamics and beyond}},
  author={Succi, S.},
  year={2001},
  publisher={Oxford university press},
  address   = {Oxford}
}

@book{Xu2022,
  title={{Complex Media Kinetics}},
  author={Xu, A. G. and Zhang, Y. D.},
  year={2022},
  publisher={Science Press},
  address   = {Beijing}
}

@article{Xu2024FOP,
  title={{Advances in the kinetics of heat and mass transfer in near-continuous complex flows}},
  author={Xu, A. G. and Zhang, D. J. and Gan, Y. B.},
  journal={Front. Phys.},
  volume={19},
  number={4},
  pages={42500},
  year={2024}
}

@book{Mccoy2009,
  title={{Advanced statistical mechanics}},
  author={McCoy, B. M.},
  year={2009},
  publisher={Oxford university press},
  address   = {Oxford}
}

@article{Lee1952PR,
  title={{Statistical theory of equations of state and phase transitions. II. Lattice gas and Ising model}},
  author={Lee, T. D. and Yang, C. N.},
  journal={Phys. Rev.},
  volume={87},
  number={3},
  pages={410},
  year={1952}
}

@book{He2009,
  title={{Lattice Boltzmann method: theory and applications}},
  author={He, Y. L. and Wang, Y. and Li, Q.},
  year={2009},
  publisher={Science Press},
  address  = {Beijing}
}

@book{Huang2015,
  title={{Multiphase lattice Boltzmann methods: Theory and application}},
  author={Huang, H. B. and Sukop, M. C. and Lu, X. Y.},
  year={2015},
  publisher={John Wiley \& Sons}
}

@article{GanYB2022JFM,
  title={{Discrete Boltzmann multi-scale modelling of non-equilibrium multiphase flows}},
  author={Gan, Y. B. and Xu, A. G. and Lai, H. L. and Li, W. and Sun, G. L. and Succi, S.},
  journal={J. Fluid Mech.},
  volume={951},
  pages={A8},
  year={2022}
}

@article{WangSE2023CP,
  title={{High-order modeling of multiphase flows: Based on discrete Boltzmann method}},
  author={Wang, S. E. and Lin, C. D. and Yan, W. W. and Su, X. L. and Yang, L. C.},
  journal={Comput. Fluids},
  volume={265},
  pages={106009},
  year={2023}
}

@article{HuangWH2025ATE,
  title={{Discrete Boltzmann method with chemical reactive mechanism for reacting flows}},
  author={Huang, W. H. and Lin, C. D. and Su, X. L. and Li, J.},
  journal={Appl. Therm. Eng.},
  volume={281},
  pages={128524},
  year={2025}
}

@article{WuQB2025CF,
  title={{Burnett-level multi-relaxation-time central-moment discrete Boltzmann modeling of reactive flows}},
  author={Wu, Q. B. and Lin, C. D. and Lai, H. L.},
  journal={Combust. Flame},
  volume={282},
  pages={114481},
  year={2025}
}

@article{LiuZP2023JMES,
  title={{Discrete Boltzmann modeling of plasma shock wave}},
  author={Liu, Z. P. and Song, J. H. and Xu, A. G. and Zhang, Y. D. and Xie, K.},
  journal={J. Mech. Eng. Sci.},
  volume={237},
  number={11},
  pages={2532-2548},
  year={2023}
}

@article{SongJH2024POF,
  title={{Plasma kinetics: Discrete Boltzmann modeling and Richtmyer--Meshkov instability}},
  author={Song, J. H. and Xu, A. G.  and Miao, L. and Chen, F. and Liu, Z. P. and Wang, L. F. and Wang, N. F. and Hou, X.},
  journal={Phys. Fluids},
  volume={36},
  number={016107},
  year={2024}
}

@article{GuoQH2025POF,
  title={{Thermodynamic nonequilibrium effects in three-dimensional high-speed compressible flows: Multiscale modeling and simulation via the discrete Boltzmann method}},
  author={Guo, Q. H. and Gan, Y. B. and Yang, B. and Wu, Y. H. and Lai, H. L. and Xu, A. G.},
  journal={Phys. Fluids},
  volume={37},
  number={4},
  year={2025}
}

@article{LinCD2019PRE,
  title={{Discrete Boltzmann modeling of unsteady reactive flows with nonequilibrium effects}},
  author={Lin, C. D. and Luo, K. H.},
  journal={Phys. Rev. E},
  volume={99},
  number={1},
  pages={012142},
  year={2019},
  publisher={APS}
}

@article{ChenF2018POF,
  title={{Collaboration and competition between Richtmyer--Meshkov instability and Rayleigh--Taylor instability}},
  author={Chen, F. and Xu, A. G. and Zhang, G. C.},
  journal={Phys. Fluids},
  volume={30},
  number={10},
  pages={102105},
  year={2018}
}

@article{LaiHL2024CF,
  title={{Investigation of effects of initial interface conditions on the two-dimensional single-mode compressible Rayleigh--Taylor instability: Based on the discrete Boltzmann method}},
  author={Lai, H. L. and Li, D. M. and Lin, C. D. and Chen, L. and Ye, H. Y. and Zhu, J.  J.},
  journal={Comput. Fluids},
  volume={277},
  number={106289},
  year={2024}
}

@article{LiYF2022FOP,
  title={{Influence of the tangential velocity on the compressible Kelvin--Helmholtz instability with nonequilibrium effects}},
  author={Li, Y. F. and Lai, H. L. and Lin, C. D. and Li, D. M.},
  journal={Front. Phys.},
  volume={17},
  number={6},
  pages={63500},
  year={2022}
}

@article{LaiHL2025POF,
  title={{Compressible Kelvin--Helmholtz instability with nonequilibrium effect under external force}},
  author={Lai, H. L. and Li, Y. F. and Lin, C. D.},
  journal={Phys. Fluids},
  volume={37},
  number={076138},
  year={2025}
}

@article{LiYF2024POF,
  title={{Kinetic investigation of Kelvin--Helmholtz instability with nonequilibrium effects in a force field}},
  author={Li, Y. F. and Lin, C. D.},
  journal={Phys. Fluids},
  volume={36},
  number={11},
  year={2024}
}

@article{LinCD2025CMA,
  title={{A flux solver based on discrete Boltzmann method for compressible flows with nonequilibrium effects}},
  author={Lin, C. D.  and Shu, C.},
  journal={Comput. Math. Appl.},
  volume={195},
  pages={14--27},
  year={2025}
}

@article{Marchuk1968AM,
  title={{Some application of splitting-up methods to the solution of mathematical physics problems}},
  author={Marchuk, G. I.},
  journal={Aplikace matematiky},
  volume={13},
  number={2},
  pages={103--132},
  year={1968}
}

@article{Godunov1959MS,
  title={{A difference scheme for numerical computation of discontinuous solutions of hydrodynamic equations}},
  author={Godunov, S. K.},
  journal={Math. Sbornik},
  volume={47},
  pages={271},
  year={1959}
}

@article{Strang1968SIAM,
  title={{On the construction and comparison of difference schemes}},
  author={Strang, G.},
  journal={SIAM J. Numer. Anal.},
  volume={5},
  number={3},
  pages={506--517},
  year={1968}
}

@book{LeVeque2002,
  title={{Finite volume methods for hyperbolic problems}},
  author={LeVeque, R. J.},
  year={2002},
  publisher={Cambridge university press}
}

@article{Dellar2011PRE,
  title={{Isotropy of three-dimensional quantum lattice Boltzmann schemes}},
  author={Dellar, P. J. and Lapitski, D. and Palpacelli, S. and Succi, S.},
  journal={Phys. Rev. E},
  volume={83},
  number={4},
  pages={046706},
  year={2011},
}

@article{Yan2013FOP,
  title={{Lattice Boltzmann model for combustion and detonation}},
  author={Yan, B. and Xu, A. G. and Zhang, G. C. and Ying, Y. J. and Li, H.},
  journal={Front. Phys.},
  volume={8},
  number={1},
  pages={94--110},
  year={2013},
}

@article{Hajabdollahi2018PRE,
  title={{Symmetrized operator split schemes for force and source modeling in cascaded lattice Boltzmann methods for flow and scalar transport}},
  author={Hajabdollahi, F. and Premnath, K. N.},
  journal={Phys. Rev. E},
  volume={97},
  number={6},
  pages={063303},
  year={2018}
}

@article{LinCD2014PRE,
  title={{Polar-coordinate lattice Boltzmann modeling of compressible flows}},
  author={Lin, C. D. and Xu, A. G. and Zhang, G. C. and Li, Y. J. and Succi, S.},
  journal={Phys. Rev. E},
  volume={89},
  number={1},
  pages={013307},
  year={2014}
}

@article{LinCD2022AAS,
  title={{Simplified two-dimensional discrete Boltzmann model of high-speed compressible reactive flows}},
  author={Lin, C. D},
  journal={Acta Aerodyn. Sin.},
  volume={40},
  number={3},
  pages={98-108},
  year={2022}
}

@article{LinCD2023CPB,
  title={{A discrete Boltzmann model with symmetric velocity discretization for compressible flow}},
  author={Lin, C. D. and Sun, X. P. and Su, X. L. and Lai, H. L. and Fang, X.},
  journal={Chinese Phys. B},
  volume={32},
  number={11},
  pages={110503},
  year={2023}
}

@article{ZhangHX1991AAM,
  title={NND schemes and their applications to numerical simulation of two-and three-dimensional flows},
  author={Zhang, H. X. and Zhuang, F. G.},
  journal={Adv. Appl. Mech.},
  volume={29},
  pages={193--256},
  year={1991}
}

\end{small}
\end{CJK*}
\bibliographystyle{ieeetr}
\end{document}